\documentstyle[12pt]{article}
\renewcommand\baselinestretch{1.15}
\advance\textheight by \topskip
\begin{document}
\def\lsim{\mathrel{\lower2.5pt\vbox{\lineskip=0pt\baselineskip=0pt
\hbox{$<$}\hbox{$\sim$}}}}
\def\gsim{\mathrel{\lower2.5pt\vbox{\lineskip=0pt\baselineskip=0pt
\hbox{$>$}\hbox{$\sim$}}}}
\def\gs{SU(2)_{\rm L} \times U(1)_{\rm Y}}
\def\wh{\hat{W}}
\def\bh{\hat{B}}
\def\wtu{\hat{W}^{\mu \nu}}
\def\wtd{\hat{W}_{\mu \nu}}
\def\btu{\hat{B}^{\mu \nu}}
\def\btd{\hat{B}_{\mu \nu}}
\def\cl{{\cal L}}
\def\nll{\cl_{\rm NL}}
\def\ecl{\cl_{\rm EChL}}
\def\fpnl{\cl_{\rm FP}^{\rm NL}}
\def\fpl{\cl_{\rm FP}}
\def\msb{{\overline{\rm MS}}}
\def\mh{M_H}
\begin{titlepage}
\title{ The electroweak chiral parameters for a heavy Higgs
in the Standard Model\thanks{Contribution to the 27$^{th}$
International Conference on High Energy Physics, Glasgow, 20-27 July
1994.}}
\author{Maria J.
Herrero\thanks{e--mail:herrero@vm1.sdi.uam.es}\ \ \ and \
\ Ester Ruiz Morales\thanks{e--mail:
meruiz@vm1.sdi.uam.es} \\[3mm] Departamento de F\'{\i}sica
Te\'orica\\ Universidad Aut\'onoma de Madrid\\ Cantoblanco,\ \ 28049
-- Madrid,\ \ Spain} \date{}
\maketitle
\def\baselinestretch{1.3}
\begin{abstract}
\noindent
 We study the electroweak interactions within the standard
electroweak theory in the case where the Higgs particle is heavy,
namely, $M_H \leq 1\;TeV$. By integrating out the Higgs boson to one
loop we find the complete effective lagrangian, called electroweak
chiral lagrangian, that is $SU(2)_L\times U(1)_Y$ invariant and
contains the whole set of operators up to dimension four. The values
of the chiral parameters representing the non-decoupling effects of
the heavy Higgs are presented. Some examples have been chosen to show
the applicability of this effective lagrangian approach to compute
phenomenologically relevant quantities that either are being
meassured at present or will be meassured in future experiments.
\end{abstract}

\vskip-20cm
\rightline{{\bf \large FTUAM 94/12}}
\rightline{{\bf \large May 1994}}
\vskip-12mm
\leftline{{\bf \large ICHEP94 Ref. 0303}}
\leftline{{\large Submitted to Pa-01, Pl-03}}

\end{titlepage}

\newpage

\def\baselinestretch{1.25}
\section{Introduction}

The electroweak chiral Lagrangian\cite{AB,L} (EChL) provides the most
general parametrization for the spontaneous breaking of the $\gs$
symmetry if the would-be Goldstone bosons (GB) are the only modes of
the symmetry breaking sector to be considered at low energies.  The
GB fields are parametrized non-linearly such that the electroweak
chiral Lagrangian built up with these modes and the gauge fields is
manifestly $\gs$ invariant.  The price to be paid in this Higgs-less
parametrization is that the resulting low energy theory is
non-renormalizable, and a tower of effective operators of increasing
dimension has to be added at each loop order to render the theory
finite.

{}From the point of view of effective field theories, the electroweak
chiral Lagrangian can be regarded as the low energy limit of an
underlying fundamental theory, where some heavy fields have been
integrated out inducing additional higher dimension operators. These
effective operators can, in principle, be determined if the
underlying fundamental interactions are known. In a perturbative
approach, it is done by explicit calculation of the relevant loop
diagrams and by matching the predictions of the full underlying
theory (in which heavy particles are present) and those of the low
energy effective theory (with only light degrees of freedom) at some
reference scale \cite{G}. The matching procedure fixes the value of
the effective parameters in terms of the parameters of the full
theory, ensuring the equality of the two theories at low energy. This
combined picture of integrating out the heavy fields and matching the
predictions of the two theories has been recently applied to some
particular situations in electroweak interactions
\cite{FMM,HR1,ARC}.

We have derived\cite{HR1,HR2} the EChL that parametrizes electroweak
interactions when the underlying theory is the standard model with a
heavy Higgs.  We deduce the values of the EChL parameters by
integrating out the Higgs field to one loop  and by matching the
standard model predictions in the large $\mh$ limit with the
predictions from the chiral Lagrangian to one loop order.  By large
$\mh$ limit we mean the situation where the mass of the Higgs is much
larger than the available external momenta (p$^2 \ll
\mh^2$) and other particle masses (m$^2 \ll \mh^2$, m= M$_{\rm W}$,
M$_{\rm Z}$), but not so high that perturbation theory is unreliable
($\mh^2 \lsim 1 $TeV$^2$).

In section 2 we briefly present the EChL and discuss the
renormalization of the effective theory. In section 3, we describe
the matching of the standard model with a heavy Higgs and the EChL,
and give the corresponding values of the chiral parameters.  Section
4 is devoted to discuss some examples of how the EChL formalism can
be applied to calculate electroweak radiative corrections to
different phenomenologically relevant quantities.

\section{The electroweak chiral Lagrangian}

The electroweak chiral Lagrangian based on the symmetry breaking
pattern $\gs \rightarrow U(1)_{em}$ is written in terms of the gauge
and GB fields \cite{AB,L}
\begin{equation}
\ecl = \cl_{\rm   NL} + \sum_{i=0}^{13} \cl_{i} ,\label{ECL}
\end{equation}
where
\begin{equation}
 \nll  =  \frac{v^2}{4}\; Tr\left[ D_\mu U^\dagger D^\mu U \right] +
\frac{1}{2}\; Tr\left[ \wtd \wtu + \btd \btu
\right] + \cl_{\rm R_\xi} + \fpnl
\label{NLL}
\end{equation}
is the Lagrangian of the gauged non-linear sigma model and the set of
$\gs$ and CP-invariant\footnote{ There is another CP conserving but C
and P violating operator $\cl_{14}$  \cite{AW}.  It is cero in
absence of fermionic loop contributions and will not be considered
here.} operators up to dimension four\footnote{In a generic R$_\xi$-
gauge, the BRS invariance of the lagrangian implies the existence of
operators involving also the ghost fields. These operators, however,
are nor needed in any calculation of physical quantities at one loop,
and therefore we will not include them.} can be parametrized as
\begin{equation}
\begin{array}{ll}
\cl_{0}  =  a_0 \frac{v^2}{4} \left[ Tr\left( T V_\mu \right)
\right]^2 \hspace{1cm} &
\cl_{7}  =  a_7 Tr\left( V_\mu V^\mu \right) \left[ Tr\left( T V^\nu
\right) \right]^2\\[2mm]
\cl_{1}  =  a_1 \frac{i g g'}{2} B_{\mu\nu}  Tr\left( T \wtu
\right)  &
\cl_{8}  =  a_8  \frac{g^2}{4} \left[ Tr\left( T \wtd \right)
\right]^2 \\[2mm]
\cl_{2}  =  a_2 \frac{i g'}{2} B_{\mu\nu} Tr\left( T [V^\mu,V^\nu ]
\right) \hspace{1cm} &
\cl_{9}  =  a_9  \frac{g}{2} Tr\left( T \wtd \right) Tr\left(
T [V^\mu,V^\nu ] \right) \\[2mm]
\cl_{3}  =  a_3  g Tr\left( \wtd [V^\mu,V^\nu ]\right)
\hspace{1cm} &
\cl_{10}  =  a_{10} \left[ Tr\left( T V_\mu \right) Tr\left( T V_\nu
\right) \right]^2 \\[2mm]
\cl_{4}  =  a_4  \left[ Tr\left( V_\mu V_\nu \right) \right]^2
\hspace{1cm} &
\cl_{11}  =  a_{11} Tr\left( ( D_\mu V^\mu )^2 \right)\\[2mm]
\cl_{5}  =  a_5  \left[ Tr\left( V_\mu V^\mu \right) \right]^2 &
\cl_{12}  =  a_{12} Tr\left( T D_\mu D_\nu V^\nu \right) Tr
\left( T V^\mu \right)\\[2mm]
\cl_{6}  =  a_6 Tr\left( V_\mu V_\nu \right) Tr\left( T V^\mu
\right) Tr\left( T V^\nu \right) \hspace{1cm} &
\cl_{13} =  a_{13} \frac{1}{2} \left[ Tr \left( T D_\mu V_\nu
\right) \right]^2 \label{Li}
\end{array}
\end{equation}
\vspace{-1cm}

\begin{center}$T \equiv  U \tau^3 U^\dagger, \hspace{2cm} V_\mu
\equiv (D_\mu U) U^\dagger.$ \end{center}
where $U = \displaystyle{\exp\left( {i \;
\frac{\vec{\tau}\cdot\vec{\pi}}{v}}\right)}$ is the GB
unitary matrix and $\wtd, \btd$ are the field strengths for the
$SU(2)_{\rm L}$ and $U(1)_{\rm Y}$ gauge fields $\wh_\mu =  -i
\vec{W}_\mu \cdot \vec{\tau}/2 ,
\bh_\mu = -i B_\mu \; \tau^3 / 2$.

Although this parametrization of electroweak interactions is
non-renormalizable in the usual sense, this by no means imply that
the predictive power is lost.  The electroweak chiral Lagrangian can
be properly renormalized at one loop, and all the infinities can be
absorved into redefinitions of the parameters of the lagrangian
\begin{center}$\begin{array}{ll}
B_{\mu}^b  \; = \; \widehat{Z}_B^{1/2} \; B_\mu, \hspace{1cm}& g'^b
\; =\; \widehat{Z}_B^{-1/2}\; ( g' - \widehat{\delta g'} ), \\[2mm]
\vec{W}_\mu^b \; =\;  \widehat{Z}_W^{1/2}\; \vec{W}_\mu, &
 g^b \; =\; \widehat{Z}_W^{-1/2}\; ( g - \widehat{\delta g} ),
\\[2mm]
\vec{\pi}^b \; =\; \widehat{Z}_\pi^{1/2}\; \vec{\pi}, &
v^b \; = \; \widehat{Z}_\pi^{1/2}\; ( v -\widehat{ \delta v}),\\[2mm]
\xi_B^b \; = \; \xi_B\; ( 1 + \widehat{\delta \xi}_B), \hspace{1cm}&
\xi_W^b \; = \; \xi_W \;( 1 + \widehat{\delta \xi}_W),
\end{array}$
\begin{equation}
a_i^b \; = \; a_i(\mu) \; + \; \delta a_i , \label{rep}
\end{equation}
\end{center}
where $ \widehat{Z}_i \equiv 1 + \widehat{\delta Z_i}$ and the
superscript b denotes bare quantities.

Once a particular renormalization prescription is chosen to fix the
counterterms in the effective theory, the renormalized parameters in
the right hand side of eq.(\ref{rep}) -that are in general
$\mu$-scale and renormalization prescription dependent- remain as
free parameters that can not be determined within the framework of
the low energy theory.  The values of the renormalized chiral
parameters $a_i(\mu)$ can be constrained from experiment as they are
directly related to different observables in scattering
processes\cite{SCT} and in precision electroweak
measurements\cite{EPM}; but to have any theoretical insight on their
values, one has to choose a particular model for the dynamics of the
symmetry breaking sector.

\section{Chiral parameters for a heavy Higgs}

If the underlying fundamental theory is the standard model(SM) with a
heavy Higgs, the values of the chiral parameters can be  determined
by matching the predictions of the standard model in the large Higgs
mass limit and those of the EChL at one loop level\cite{HR1,HR2}.

We have imposed the strongest form of matching by requiring that all
renormalized one-light-particle irreducible (1LPI) Green's functions
are the same in both theories at scales $\mu \leq \mh$.  This
matching condition is equivalent to the equality of the light
particle effective action in the two descriptions.  In practice, it
is enough to analyse the two, three, and four-point Green's functions
with external gauge fields to fix completely the chiral parameters in
terms of the parameters of the SM.  We have worked in a general
R$_\xi$-gauge to show that the chiral parameters $a_i$ are
$\xi$-independent, and used dimensional regularization to regulate
the divergent integrals.

The SM Green's functions are non-local as they depend on $p / \mh$
through the virtual Higgs propagators.  When doing the low energy
expansion, care must be taken since clearly the operations of making
loop integrals and taking the large $\mh$ limit do not commute. Thus,
one must first regulate the loop integrals by dimensional
regularization, then perform the renormalization with some fixed
prescription and at the end take the large $\mh$ limit, with $\mh$
being the renormalized Higgs mass. We have renormalized in the on-shell
scheme.  From the computational point of view, in the large $\mh$
limit we have neglected  contributions that depend on (p$/\mh)^2$
and/or (m$/\mh)^2$ and vanish when the formal $\mh \rightarrow
\infty$ limit is taken.

The matching procedure can be summarized by the following relation
among renormalized Green's functions
\begin{equation}
\Gamma_{\rm R}^{\rm     SM}(\mu ) \; = \; \Gamma_{\rm R}^{\rm EChL}
(\mu )\; , \;\;\;\;\;\;\;\;\;\mu\leq \mh,
\end{equation}
where the large Higgs mass limit in the left-hand side must be
understood throughout.  This equation represents symbolically a
system of tensorial coupled equations (as many as 1LPI functions for
external gauge fields) with several unknowns, namely the complete set
of parameters $a_i(\mu)$ and counterterms that we are interested in
determining.  There is just one compatible solution given by the
following values of the bare electroweak chiral parameters
\begin{eqnarray}
a_0^b &  =  &  g'^2 \frac{1}{16 \pi^2} \frac{3}{8}
\left( \Delta_\epsilon - \log \frac{\mh^2}{\mu^2} +
\frac{5}{6}\right), \nonumber \\
a_1^b & = &  \frac{1}{16 \pi^2} \frac{1}{12}
\left( \Delta_\epsilon - \log \frac{\mh^2}{\mu^2}
+ \frac{5}{6} \right), \nonumber \\ a_2^b & = &  \frac{1}{16 \pi^2}
\frac{1}{24}
\left( \Delta_\epsilon - \log \frac{\mh^2}{\mu^2} +
\frac{17}{6} \right), \nonumber \\
a_3^b & = & \frac{-1}{16 \pi^2} \frac{1}{24}
\left( \Delta_\epsilon - \log \frac{\mh^2}{\mu^2} +
\frac{17}{6} \right), \nonumber \\
a_4^b & = & \frac{-1}{16 \pi^2} \frac{1}{12}
\left( \Delta_\epsilon - \log \frac{\mh^2}{\mu^2} +
\frac{17}{6}\right), \nonumber \\
a_5^b & = & \frac{v^2}{8 \mh^2} - \frac{1}{16 \pi^2} \frac{1}{24}
\left( \Delta_\epsilon - \log \frac{\mh^2}{\mu^2} + \frac{79}{3}
- \frac{ 27 \pi}{2 \sqrt{ 3}} \right), \nonumber\\ a_{11}^b  & = &
\frac{-1}{16 \pi^2}\frac{1}{24}, \nonumber \\[2mm]
a_6^b & = & a_7^b \; =\; a_8^b \; = \; a_9^b \; = \; a_{10}^b \; = \;
a_{12}^b \; = \; a_{13}^b \; = \; 0, \label{aMH}
\end{eqnarray}
where
\begin{equation}
\Delta_\epsilon  \equiv  \frac{2}{\epsilon} - \gamma_E +
\log 4 \pi.
\end{equation}
We would like to make some remarks on the previous result:
\begin{enumerate}
\item First of all, we agree with the $1/\epsilon$ dependence
of the $a_i^b$ parameters that was first calculated by Longhitano
\cite{L} looking at the divercences of the non-linear sigma model.
We see therefore that the divergences generated with the $\nll$ to
one loop are exactly canceled by the $1/\epsilon$ terms in the
$a_i^b$'s.

What is important to realize is that the matching procedure fixes
completely the values of the bare parameters $a_i^b$ in terms of the
renormalized parameters of the SM.  Once we have chosen a particular
substraction scheme to fix the counterterms of the EChL, in our case
\begin{equation}
\begin{array}{ll}
\delta a_0 \; = \; {\displaystyle g'^2 \frac{1}{16 \pi^2} \frac{3}{8}
\Delta_\epsilon}, \hspace{1cm} &
\delta a_3 \; = \; {\displaystyle \frac{-1}{16 \pi^2} \frac{1}{24}
\Delta_\epsilon}, \\[2mm]
\delta a_1 \; = \;{\displaystyle \frac{1}{16 \pi^2} \frac{1}{12}
\Delta_\epsilon},  &
\delta a_4 \; = \; {\displaystyle \frac{ - 1}{16 \pi^2} \frac{1}{12}
\Delta_\epsilon},  \\[2mm]
\delta a_2 \; = \;{\displaystyle \frac{1}{16 \pi^2} \frac{1}{24}
\Delta_\epsilon},  &
\delta a_5 \; = \; {\displaystyle \frac{- 1}{16 \pi^2} \frac{1}{24}
\Delta_\epsilon },
\end{array} \label{CT}
\end{equation}
then the  corresponding renormalized values of the chiral parameters
are given in terms of the SM parameters as follows:
\begin{equation}
\begin{array}{ll}
a_0 (\mu)  =  {\displaystyle g'^2  \frac{1}{16 \pi^2}
\frac{3}{8}\left( \frac{5}{6} - \log\frac{M_H^2}{\mu^2} \right)},&
\hspace{-1cm}a_3 (\mu)  =  {\displaystyle \frac{-1}{16 \pi^2}
\frac{1}{24}
\left( \frac{17}{6} - \log\frac{M_H^2}{\mu^2} \right)},\\[4mm]
a_1 (\mu)  =  {\displaystyle \frac{1}{16 \pi^2}  \frac{1}{12}
\left( \frac{5}{6} - \log\frac{M_H^2}{\mu^2} \right)}, &
\hspace{-1cm}a_4 (\mu)  =  {\displaystyle \frac{- 1}{16 \pi^2}
\frac{1}{12}
\left( \frac{17}{6} - \log\frac{M_H^2}{\mu^2} \right)},\\[4mm]
a_2 (\mu)  =  {\displaystyle \frac{1}{16 \pi^2} \frac{1}{24}
\left( \frac{17}{6} - \log\frac{M_H^2}{\mu^2} \right)}, &
\hspace{-1cm}a_{11} (\mu)  = {\displaystyle
\frac{-1}{16 \pi^2}\frac{1}{24}},\\[4mm]
a_5 (\mu)  =  {\displaystyle \frac{v^2}{8 \mh^2} -
 \frac{1}{16 \pi^2}  \frac{1}{24}
\left( \frac{79}{3} - \frac{ 27 \pi}{2 \sqrt{3}}
 - \log\frac{M_H^2}{\mu^2} \right)}, \\[4mm]
a_i (\mu) = 0, \hspace{3mm} i = 6,7,8,9,10,12,13.
\end{array} \label{aR}
\end{equation}
\item The divergences and the logarithmic running with
the scale $\mu$ depend only on the $\gs$ symmetry
and therefore, they are the same for any
underlying dynamics for the symmetry breaking sector that
has the EChL as its low energy effective theory.
\item Eqs.(\ref{aMH}) give the complete non-decoupling effects
of a heavy Higgs, that is, the leading logarithmic dependence
on $\mh$ and the next to leading constant contribution
to the electroweak chiral parameters.
The $a_i$'s are accurate up to corrections of the order
$(p^2/\mh^2)$ where $p \approx M_Z$ and higher order
corrections in the loop expansion.
\item We demonstrate that the $a_i$'s are gauge independent, as
expected.
\item  There is only one effective operator $a_5$ that gets
a tree level contribution.
It's value depends on the renormalization prescription that one has
chosen in the standard model, on-shell in our case.
\item Only one custodial breaking operator $a_0$, which has dimension
2, is generated when integrating out the Higgs at one loop.  No
custodial breaking operator of dimension four are generated.
\end{enumerate}

We believe that this effective field theory calculation provides
valuable information since it clarifies the relation between the
linear and non-linear approach to electroweak interactions.  The
chiral parameters of eqs.(\ref{aMH}) also serve as reference values
to be compared with the corresponding predictions from other possible
alternatives for the symmetry breaking.

\section{ Calculating observables with the EChL}

In this section we will show, as an example, the explicit calculation
of the radiative corrections to $\Delta r$ within the electroweak
chiral Lagrangian approach.  In the on-shell scheme, $\Delta r$ is
defined as\cite{CH}
\begin{eqnarray}
\Delta r  & = & \Sigma'_{\gamma \gamma}(0) +
\frac{c_w^2}{s_w^2} \left( \frac{\Sigma_{WW}(M_W^2)}{M_W^2}
-  \frac{\Sigma_{ZZ}(M_Z^2)}{M_Z^2} + \frac{2 s_W}{c_W}
\frac{\Sigma_{\gamma Z}(0)}{M_Z^2} \right)\nonumber\\ & & +
\frac{\Sigma_{WW}(0) - \Sigma_{WW}(M_W^2)}{M_W^2} + \frac{\alpha}{4
\pi s_W^2} \left( 6 + \frac{7 - 4 s_W^2} {2 s_W^2} \log c_W^2 \right)
\end{eqnarray}
where $\Sigma$ denote unrenormalized self-energies, and therefore
will have the contribution from the loop diagrams generated with
$\nll$ and contribution from the bare $a_i^b$ effective parameters.
Considering only bosonic contributions, the loop diagrams
give\footnote{See the second paper in ref.\cite{EPM}}
\begin{equation}
\Delta r |^{\rm bosonic}_{\rm loops} = \frac{g^2}{16 \pi^2} \left[
\frac{11}{12} \Delta_\epsilon - \frac{11}{12}
\log \frac{M_W^2}{\mu^2} + a(M_W^2,M_Z^2) \right]
\end{equation}
where $a(M_W^2,M_Z^2)$ include the finite and scale-independent
contribution from the EChL loops to $\Delta r$
\begin{eqnarray}
a(M_W^2,M_Z^2) & = & \log c_W^2 \left( \frac{5}{c_W^2} -1 +
\frac{3 c_W^2}{s_W^2} - \frac{17}{4 s_W^2 c_W^2} \right)
- s_W^2 (3 + 4 c_W^2) F(M_Z^2,M_W,M_W)\nonumber \\ & & + I_2(c_W^2)
(1 - \frac{c_W^2}{s_W^2}) + \frac{c_W^2}{s_W^2} I_1(c_W^2) +
\frac{1}{8 c_W^2} ( 43 s_W^2 - 38) \nonumber \\
& & + \frac{1}{18} (154 s_W^2 - 166 c_W^2) + \frac{1}{4 c_W^2} +
\frac{1}{6} +
\left( 6 + \frac{7 - 4 s_W^2} {2 s_W^2} \log c_W^2 \right).
\end{eqnarray}

The contribution to $\Delta r$ comming from the effective operators
can be easily obtained from their corresponding contribution to the
self energies given in \cite{HR1}
\begin{equation}
\Delta r |_{{\rm a}_i} \; = \; - 2 g^2 a_1^b - 2 \frac{c_W^2}{s_W^2}
a_0^b \; = \; - 2 g^2 a_1(\mu) - 2 \frac{c_W^2}{s_W^2} a_0(\mu) -
\frac{g^2}{16 \pi^2} \frac{11}{12} \Delta_\epsilon,
\end{equation}
where we have taken the substraction scheme defined in eqs.(\ref{CT})
for the effective chiral parameters.

Finally, the expression for $\Delta r$ in the EChL formalism is
\begin{equation}
\Delta r ^{\rm EChL} \; = \; - 2 g^2 a_1(\mu) - 2 \frac{c_W^2}
{s_W^2} a_0(\mu) - \frac{g^2}{16 \pi^2} \frac{11}{12}
\log \frac{M_W^2}{\mu^2} +
\frac{g^2}{16 \pi^2} a(M_W^2,M_Z^2).
\end{equation}
The divergences of the bosonic loop contributions have been cancelled
by the $\delta a_0$ and $\delta a_1$ counterterms. The $\mu$-scale
dependence of the loop contributions is also properly cancelled by
the scale dependence of the renormalized $a_i(\mu)$'s, so that the
observable is $\mu$-scale and renormalization prescription
independent.

In the particular case of the SM with a heavy Higgs, one has just to
substitute the values of $a_i(\mu)$ from eqs.(\ref{aR}) to obtain
\begin{equation}
\Delta r ^{\rm heavy}_{\rm Higgs} =
\frac{g^2}{16 \pi^2}\left( \frac{11}{12} \log \frac{\mh^2}{M_W^2}
- \frac{5}{6} \right) +
\frac{g^2}{16 \pi^2} a(M_W^2,M_Z^2).
\end{equation}
which agrees with the result given in \cite{CH}.

One can similarly obtain the heavy Higgs contributions to the rest of
electroweak parameters\cite{HR1,HR2}.

\end{document}